\documentclass[final,5p,times,sort,compress]{elsarticle}

\usepackage{amssymb}
\usepackage{amsmath}

\usepackage{amsfonts}
\usepackage{hyperref}
\usepackage{color,graphicx} 
\usepackage{subfigure}
\usepackage{dcolumn}
\usepackage{booktabs}
\usepackage{geometry}
\usepackage{multirow}
\usepackage{array}
\usepackage{bm}
\usepackage{comment}
\usepackage{xcolor}
\usepackage{appendix}

\usepackage{float}
\begin{document}

\begin{frontmatter}



\title{Probing the magnetic ground state and magnetoelastic coupling in  double perovskite ruthenate : Ca$_2$ScRuO$_6$} 


\author[label1]{Asha Ann Abraham} 
\author[label2,label3]{Anjali Kumari}
\author[label4]{Md Aktar Hossain}
\author[label5]{Sanjoy Kr Mahatha}
\author[label4]{Saikat Das}
\author[label2,label3]{A. K. Bera}
\author[label1]{Soham Manni}
\affiliation[label1]{organization={Department of Physics},
            addressline={Indian Institute of Technology Palakkad}, 
            city={ Palakkad},
            postcode={678623}, 
            state={Kerala},
            country={India}}
\affiliation[label2]{organization={Solid State Physics Division},
            addressline={Bhabha Atomic Research Centre}, 
            city={Mumbai},
            postcode={400085}, 
            state={Maharashtra},
            country={India}}
\affiliation[label3]{organization={Homi Bhabha National Institute},
            addressline={Anushaktinagar}, 
            city={Mumbai},
            postcode={400094}, 
            state={Maharashtra},
            country={India}}
\affiliation[label4]{organization={Department of Physics},
            addressline={Indian Institute of Technology Kharagpur}, 
            city={Kharagpur},
            postcode={721302}, 
            state={West Bengal},
            country={India}}
\affiliation[label5]{organization={UGC-DAE Consortium for Scientific Research},
            addressline={University Campus}, 
            city={Indore},
            postcode={452001}, 
            state={Madhya Pradesh},
            country={India}}

\begin{abstract}
Ruthenates, materials with a single magnetic Ruthenium (Ru) atom, often display an exotic array of ground states ranging from superconductivity to altermagnetism. In this work, we investigated the magnetic ground state of a least explored member of the $4d^3$ double perovskite ruthenate series A$_{2}$ScRuO$_{6}$ (A = Ca, Sr, Ba) : Ca$_{2}$ScRuO$_{6}$. Interestingly, temperature-dependent bulk susceptibility curve shows ferrimagnetic like behavior above the magnetic ordering at around 40 K, which were corroborated by the identification of the mixed valence states, Ru$^{5+}$ and Ru$^{4+}$ via X-ray absorption spectroscopy. Structural analysis further revealed atomic-site exchange between the Ru and Sc sites which results in the Ru mixed valence states. Neutron powder diffraction measurements detected the presence of magnetic Bragg peaks at a low temperature near 4 K and a moderate magnetoelastic coupling near the ordering temperature of 40 K. However, the corresponding symmetry analysis shows a weak Type I antiferromagnetic ground state with a reduced magnetic moment of 1.1$\mu_{B}$/Ru atom.  Our findings establish an unusual magnetic ground state in the Mott insulating Ca$_{2}$ScRuO$_{6}$ where a long range ordered antiferromagnet coexists with small magnetic clusters which manifests a ferrimagnetic-like high temperature inverse magnetic susceptibility. This system presents a unique platform to study long-range magnetic order in the presence of antisite disorder. 
\end{abstract}




\end{frontmatter}




\section{Introduction}
Ruthenates are host to a diverse range of exotic phases which includes superconductivity, 2D magnetism, Mott insulating states, heavy fermi liquid behavior, altermagnetism and promising avenues for oxide spintronics technologies.\cite{Sr2RuO4,intro_AgRuO3,intro_SrRuO3,CaRuO3,intro_RuO2,spintronics_application}. The interplay between electron- electron Coulombic repulsion ($U$) and spin orbit interaction ($SOI$) often result in a plethora of electronic and magnetic ground states in ruthenates and other 4d/5d transition metal oxides (TMOs) \cite{annual_condensed_matter}. Double perovskite (DP) TMOs with the general formula A$_{2}$BB$^{'}$O$_{6}$, provide a versatile platform to systematically investigate the interplay of the competing interactions across the different diverse ground states in DP ruthenates e.g. Long-range ordered (LRO) antiferromagnetic (AFM) state, superconductivity, spin-glass behavior, incommensurate magnetic order, non-coplanar magnetic textures and ferrimagnetism.\cite{SYRO_Mott_AFM_1, SYRO_Mott_AFM_2, superconductivity_in_DP, SCRO_ferrimagnetism, CaSrFeRuO6_spinglass, La2NaRuO6_incommensurate, Ba2YRuO6_Ba2LuRuO6}.\par
DP ruthenates with a single magnetic Ru atom significantly reduces the magnetic complexity observed in mixed DP ruthenates. This enables in exploring the magnetism of 4d Ru atom taking into account the superexchange interactions, magnetic frustration, $SOI$, crystal electric field effects which helps in detailed investigations of different emergent phenomena intrinsic to 4d TMOs. Among these DP TMOs, 4d$^3$ DP TMOs with a single magnetic ion are an interesting class of correlated materials where the effect of $SOI$ is debatable. Conventionally for these materials, a classical spin only ground state of $S=3/2$ is predicted which is challenged in recent times\cite{theory_4d3_5d3}. The relativistic spin-orbit coupled $J_{eff} = 3/2$ ground state of AFM Ba$_2$YOsO$_6$ is an example of Spin-Orbit Coupled ($SOC$) Mott insulator in a d$^3$ system\cite{Ba2YOsO6}. But a complete picture of the magnetic ground state of different 4d$^3$ DP TMOs is still lacking despite investigations on a few DP TMOs. Two dimensional magnetic correlations in Sr$_2$YRuO$_6$, observation of unusual spin gap in  Ba$_2$YRuO$_6$, long-range orderedAFM magnetic ground state in Ca$_2$LaRuO$_6$ and  Ba$_2$LaRuO$_6$ show the intriguing behavior of Ru atom in the 4d$^3$ DP ruthenate system\cite{2d_magnetic_correlations_Sr2YRuO6,Ba2YRuO6_spingap,CLRO_BLRO}.\par
In this work we have explored the magnetic ground state of one of the least explored 4d$^3$ DP ruthenate Ca$_{2}$ScRuO$_{6}$. Previous works on the series A$_2$ScRuO$_6$ (A = Sr, Ba) show that they exhibit a long range ordered antiferromagnetic ground state with an insulating behavior. In this family, Sr$_2$ScRuO$_6$ crystallizes in a monoclinic $I2/m$ structure and Ba$_2$ScRuO$_6$ crystallizes in a cubic $Fm\bar{3}m$ structure. The AFM transition for Sr$_2$ScRuO$_6$ and the highly frustrated Ba$_2$ScRuO$_6$ (frustration index, $f$ $\approx$ 14) occur at approximately 60 K and 43 K respectively. Additionally a change in symmetry occurs in Sr$_2$ScRuO$_6$ from  $I2/m$ to $P2_{1}/n$ around 126 K.
Both Sr$_2$ScRuO$_6$ and Ba$_2$ScRuO$_6$, shows a Type-I ordered AFM ground state with an ordered magnetic moment of 1.97 $\mu_{B}$/Ru~atom and 2.04 $\mu_{B}$/Ru~atom respectively \cite{SSRO_BSRO,Ba2ScRuO6_DFT}. Therefore investigating the magnetic ground state of Ca$_{2}$ScRuO$_{6}$ is crucial for a comprehensive understanding of the factors controlling magnetism across the 4d$^3$ DP ruthenates A$_{2}$ScRuO$_{6}$.\par
 From powder X-ray diffraction and X-ray absorption spectroscopy we confirmed the crystal structure and the presence of disorder mediated mixed valency in Ca$_{2}$ScRuO$_{6}$. Bulk magnetization, heat capacity and resistivity measurements indicated the presence of short-range correlations in the system along with a Mott insulating behavior. But the neutron powder diffraction measurements indicates the presence of a long-range ordered AFM state with small magnetic clusters in Ca$_{2}$ScRuO$_{6}$ along with a moderate magnetoelastic coupling around the magnetic transition temperature.
\section{Experimental details}
Polycrystalline Ca$_{2}$ScRuO$_{6}$ were synthesized using the conventional solid state synthesis method. Stoichiometric quantities of CaO, Sc$_2$O$_3$ and Ru powder (all with purity $\geq99.9\%$) was mixed thoroughly into an homogeneous mixture. The mixture was heated at 650$^{\circ}C$, 850$^{\circ}C$, 1050$^{\circ}C$, 1200$^{\circ}C$ and 1400$^{\circ}C$ in a high-temperature programmable Muffle furnace, along with intermediate grinding and pelletization.\par
Room temperature powder X-ray diffraction (PXRD) was carried out using a Rigaku Smart Lab X-ray diffractometer with Cu-K$_\alpha$ radiation ($\lambda$ = 1.54 $\AA$).  Rietveld refinement of the PXRD data was done using General Structural and Analysis Software II (GSAS II)\cite{gsas}. Quantum Design Magnetic Property Measurement System (MPMS) was used for the dc and ac magnetization measurements in the temperature range of 1.8 K - 400 K. Heat capacity measurements were performed using a Quantum Design Physical Properties Measurement System (PPMS) in the temperature range of 2 -300 K at 0 kOe and 5 kOe magnetic field. Resistivity measurements were carried out on cuboidal-shaped highly pressurized pellets using Quazar Tech's XPLORE 1.2 Physical Quantities Measurement System (PQMS) in the temperature range of 80 - 300 K at zero field by applying a constant voltage of 10 V.\par
Soft X-ray absorption spectroscopy (SXAS) measurements at the Ru M$_{2,3}$-edges were performed at the SXAS beamline (BL-01) of the Indus-2 synchrotron facility, housed at the Raja Ramanna Centre for Advanced Technology (RRCAT) in Indore, India. The experiments were conducted at room temperature using total electron yield (TEY) mode with an energy resolution of about 0.5 eV and are presented here after background corrections.\par
The temperature-dependent neutron powder diffraction (NPD) measurements were done using the PD-I and PD-II diffractometers at the Dhruva research reactor, Bhabha Atomic Research Centre, Mumbai, India \cite{paranjpe1989neutron, paranjpe2002neutron}. NPD patterns were recorded over a wider angular range ($5^{\circ} < 2\theta < 140^{\circ}$) at several temperatures between 10 and 300 K using the PD-II (Ge (331) monochromator, $\lambda$ = 1.2443 $\AA$) and five linear position-sensitive detectors (PSDs). Additional NPD patterns were recorded at 4 K in the magnetic ordered state and 70 K in the paramagnetic state, using the PD-I ($\lambda$ = 1.094 $\AA$) with long counting time ($\sim$ 48 hours per pattern) to measure weak magnetic Bragg peaks. For these measurements, the powder Ca$_{2}$ScRuO$_{6}$ samples were packed into an 8 mm diameter cylindrical vanadium-can.  The vanadium-can was mounted on the cold finger of a helium-4 closed-cycle refrigerator (CCR). The neutron diffraction data were analyzed using the Rietveld refinement method using the FULLPROF suite computer program \cite{rodriguez1993recent}.\par
Temperature-dependent neutron depolarization (ND) measurements were conducted on Ca$_{2}$ScRuO$_{6}$ from 4 K to 300 K using the polarized neutron spectrometer (PNS, $\lambda$ = 1.201 $\AA$) at Dhruva reactor, BARC, Mumbai. The powdered sample was packed in a rectangular Al-can with its flat surface exposed to the neutron beam. Transmitted beam polarization measurements were performed in warming cycle under a guide magnetic field of 35 Oe to maintain neutron beam polarization.

\section{Results and Discussions}
\subsection{Structural Characterization}
\begin{figure}[ht]
    \centering
    \includegraphics[width=\linewidth]{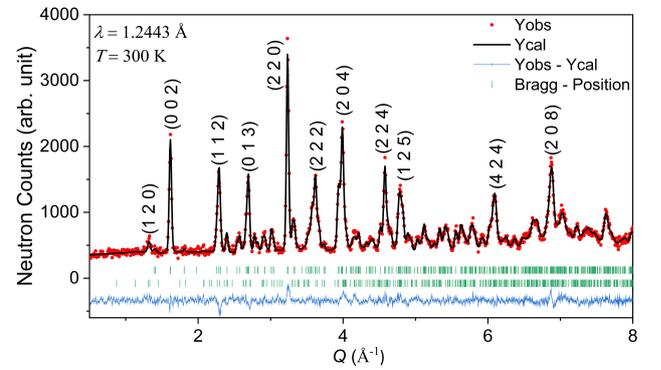}
    \caption{Neutron powder diffraction pattern of Ca$_{2}$ScRuO$_{6}$ at 300 K (PD-II Data). The observed and calculated diffraction patterns are shown by the filled circles (red) and solid lines (black), respectively. The difference between observed and calculated patterns is shown by the thin line (blue) at the bottom of the panel. The vertical bars (green) are the allowed Bragg peak positions.}
    \label{NPD_RR}
\end{figure}
\begin{table}[H]
\centering
{\footnotesize
\caption{Refined lattice parameters of Ca$_{2}$ScRuO$_{6}$ at T = 300 K obtained from NPD.}
\begin{tabular}{ccc}
\hline
Crystal System&& Monoclinic \\
Space group&& \textit{P2$_1$/n} \\
a ($\AA$)&&5.4357 (4) \\
b ($\AA$)&&5.5629(5)\\
c ($\AA$)&&7.7585(6)\\
$\beta$ &&89.92 (2)\\
$V$($\AA^3$)&& 234.61(3)\\
$R_{w}$&& 4.61 $\%$\\
\hline
\end{tabular}}
\label{RRlatticeparameters}
\end{table}
Crystal structure of Ca$_{2}$ScRuO$_{6}$ was determined from the Rietveld refinement of both NPD and lab PXRD data. FIG.\ref{NPD_RR} shows the Rietveld analysis of the NPD data at T = 300 K. The Rietveld analysis confirmed that Ca$_{2}$ScRuO$_{6}$ crystallizes in a monoclinic structure with the space group $P2_{1}/n$. Other similar systems, namely Ca$_2$XRuO$_6$ (X = Y, La) and Ca$_2$ScOsO$_6$ also crystallize in the $P2_{1}/n$ space group \cite{CLRO_BLRO, CSOO_refinement_parameters}. Phase composition analysis revealed a phase of Ca$_{2}$ScRuO$_{6}$ ($\sim$ 91 wt$\%$) with a minor secondary phase of CaSc$_{2}$O$_{4}$ ($\sim$ 9 wt$\%$). The CaSc$_{2}$O$_{4}$ phase exhibits an orthorhombic crystal symmetry (space group : $Pnma$). The refined NPD lattice parameters of Ca$_2$ScRuO$_6$ at T =300 K are given in Table \ref{RRlatticeparameters}. The corresponding Rietveld analysis of the PXRD data, crystal structure, refined lattice parameters and atomic coordinates  of Ca$_{2}$ScRuO$_{6}$ are given in FIG. \ref{RR_CSRO} and TABLE\ref{RR_neutron}, \ref{RRlatticeparameters_XRD}, \ref{wyCLRO} of \ref{app_A} respectively.\par
From the structural analysis of the Rietveld refinement it was found that ScO$_6$ octahedra ($2b$ site) forms a nearly ideal octahedra. It is characterized by three distinct but almost similar bond lengths (Sc-O1 = 2.060(12) $\AA$, Sc-O2 = 2.054(15) $\AA$, and Sc-O3 = 2.035(17) $\AA$) and a minimal variation in bond angles (O-Sc-O and O-Ru-O) as shown in Table \ref{bond_length and bond_angle} of \ref{app_A}. Similarly, RuO$_6$ octahedron ($2d$ site) is slightly distorted with three close values of bond lengths; Ru-O1 = 1.940(12) $\AA$, Ru-O2 = 1.962(14) $\AA$, and Ru-O3 = 1.991(17) $\AA$ (Table \ref{bond_length and bond_angle}). The shorter mean bond length ($\sim$1.96 $\AA$) for Ru-O as compared to the same for Sc-O ($\sim$2.05 $\AA$) is in agreement with the smaller ionic radius of Ru$^{5+}$ (0.565 $\AA$) than that of Sc$^{3+}$ (0.745 $\AA$). The alternating ScO$_6$ and RuO$_6$ octahedra are connected along all the crystallographic directions by sharing corners and the connection angles Sc-O-Ru deviates largely (149.9 $^{\circ}$ and 151.2 $^{\circ}$ within the \textit{ab} plane and 151.6 $^{\circ}$ along the \textit{c} axis) from the ideal value of 180 $^{\circ}$ which reveals the presence of significant octahedral titling in Ca$_{2}$ScRuO$_{6}$. The octahedral tilting angles (180- (Sc-O-Ru)/2) was found to be 15.05 $^{\circ}$ and 14.4 $^{\circ}$ within the \textit{ab} plane and 14.2$^{\circ}$along the \textit{c} axis.\par
Within the monoclinic crystal structure of the space group $P2_{1}/n$, there are two independent sites for the transition metal ions ($2b$ for Sc and $2d$ for Ru) along with three oxygen sites ($4e$, $4e$, and $4e$) and one Ca site ($4e$ site). Our analysis indicated a small ($\sim14\%$) intermixing of the transition metal ions Sc and Ru creating an antisite disorder. This site mixing  can possibly create an additional Ru valence state apart from Ru$^{5+}$ valence state which occurs in the absence of an intersite mixing \cite{mixedvalency_clusterspinglass}.
\subsection{XAS Analysis}
\begin{figure}[ht]
    \centering
    \includegraphics[width=\linewidth]{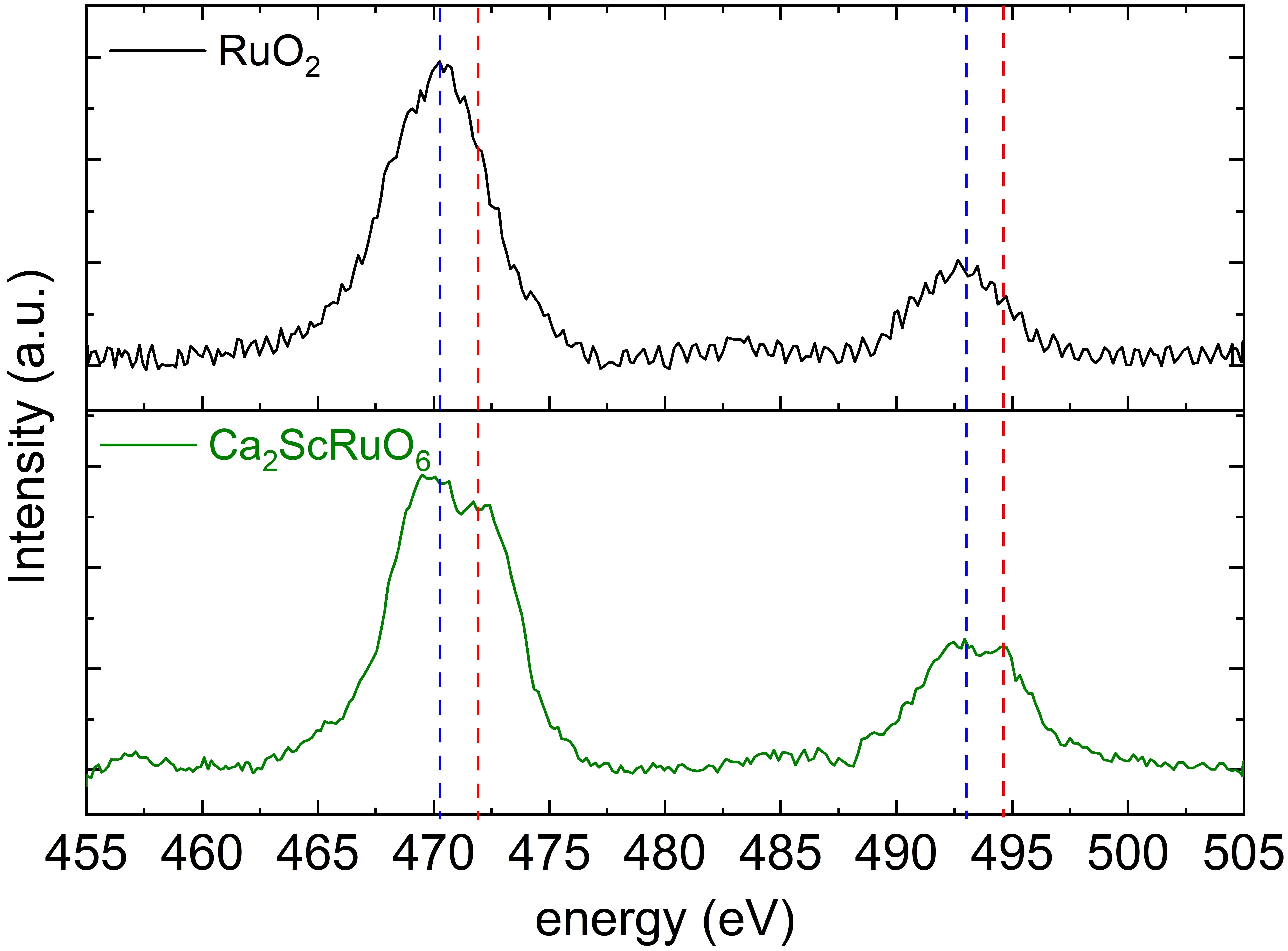}
    \caption{Background subtracted Ru M$_{2,3}$-edge XANES spectra of RuO$_2$ and Ca$_2$ScRuO$_6$ compounds measured at room temperature.}
    \label{XAS_CSRO}
\end{figure}
X-ray Absorption Spectroscopy (XAS) is a powerful synchrotron-based technique widely employed to probe the oxidation states, local coordination environments and electronic structures of materials with element-specific sensitivity. FIG.\ref{XAS_CSRO} presents the Ru M$_{2,3}$ edge X-ray absorption near-edge spectroscopy (XANES) spectra of standard RuO$_2$ and Ca$_2$ScRuO$_6$ compounds. The M$_{2,3}$ edge corresponds to the energy range where Ru 3\emph{p} core electrons are excited into unoccupied 4\emph{d} and 5\emph{s} states. The labels M$_2$ and M$_3$ arise from the spin-orbit splitting of the 3\emph{p} core levels. The spin orbit splitting of the two peaks is about 22 eV. The XANES spectra of Ca$_2$ScRuO$_6$ exhibited distinct splitting of the Ru  M$_{2,3}$ edges, with the lower-energy feature closely aligning with that of RuO$_2$, indicative of Ru$^{4+}$, while the higher-energy component corresponding to the Ru$^{5+}$ oxidation state. The spectral signature from XANES reveal a mixed-valence character of Ru in Ca$_2$ScRuO$_6$ which may have originated due to the antisite Sc/Ru disorder. This shall significantly influence the magnetic ground state of Ca$_2$ScRuO$_6$.
\subsection{Magnetization Measurements}
\begin{figure*}[t]
    \centering
    \includegraphics[width=1\linewidth]{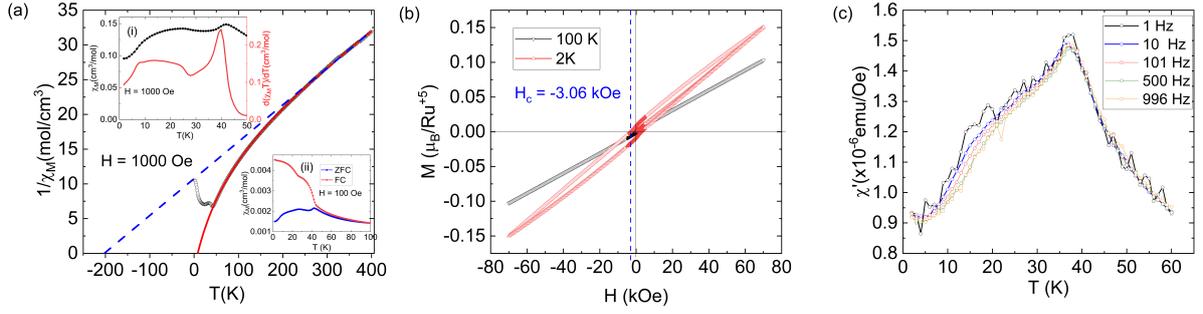}
    \caption{(a) $1 /\chi_M$ $vs.$ $T$ data of Ca$_{2}$ScRuO$_{6}$ at an external magnetic field, H = 1000 Oe (black circles). The red curve represents a  ferrimagnetic like behaviour. The dashed blue line is the paramagnetic CW fit. Inset (i) shows the $\chi _{M}$ $vs.$ $T$ and  $\frac{d(\chi _{M}T)}{dT} vs. T$ and inset (ii) shows the low field ($H = 100 Oe$) magnetization data ($\chi _{M}$  $vs.$  $T$)in ZFC and FC. (b) Field dependent magnetization ($M$ $vs.$ $H$) data for Ca$_{2}$ScRuO$_{6}$ at 2K and 100K.(c) Temperature variation of real part of ac magnetic susceptibility ($\chi^{'}$ $vs.$ $T$) at different  frequencies. }
    \label{MT_CSRO}
\end{figure*}

FIG \ref{MT_CSRO}(a) shows the temperature-dependent inverse dc molar magnetic susceptibility ($1/\chi_M$ $vs.$ $T$) in an external field ($H$) of 1000 Oe. $1/\chi _{M}$ deviates from the linear Curie-Weiss ($CW$) behavior below 250 K, rather shows a hyperbolic behavior down to the magnetic transition. High temperature non-linear inverse magnetic susceptibility is usually seen for long-range ordered  ferrimagnets, cluster spin glass and  Griffiths phase systems\cite{MT_hyperbolic_cluster_spin_glass,Griffithphase_1,Griffithphase_2,LRO_ferrimagnetism}. For magnetic systems with Griffiths phase, magnetic susceptibility follows a power law behavior which is not the case for Ca$_{2}$ScRuO$_{6}$. Moreover, Griffiths phase with large negative CW temperature is exceedingly rare.\cite{Griffith_phase_AFM}. $1/\chi_{M}(T)$ provides the best fit with ferrimagnetic susceptibility given by, $\frac{1}{\chi _{M}} = \frac{1}{\chi _{0}} + \frac{T}{C} - \frac{b}{T -\theta }$ in the temperature range of 50 - 400K (red curve, FIG. \ref{MT_CSRO}(a)). Here, $\theta$, $\chi _0$ and $b$ are functions of the intra ($n_{AA}$, $n_{BB}$) and inter ($n_{AB}$) lattice molecular field constants and $C$ is the Curie constant for a bipartite lattice with A and B sublattices which results in  a hyperbolic behavior\cite{Cullity_magnetism,LRO_ferrimagnetism}. The fitted parameters are $\theta$ = -76(2) K, $C$ = 22.4(1) cm$^3-$K/mol, $b$ = 1500(50) mol-K/cm$^3$, $\chi_0$ = 0.0579(6) mol/cm$^3$. The high temperature $1/\chi_{M}(T)$ asymptotically approaches to a linear $CW$ behavior, $\frac{1}{\chi _{M}} =  \frac{T -\theta_{cw}}{C}$ (dashed blue line), where $\theta_{cw}$ is the Curie-Weiss temperature. The fitting in the temperature range of 250 -400 K results in an effective paramagnetic moment, $\mu_{eff}$ of 3.47 $\mu_B$ and $\theta_{cw}$ of -202(1) K. The negative $\theta_{cw}$ indicates the strong antiferromagnetic correlations present in the system. This $\mu_{eff}$ is slightly reduced from the theoretical $\mu_{eff}$ of 3.87$\mu_B$ for the Ru$^{5+}$ ion which has $S = 3/2$. The presence of two magnetic ions of ruthenium, Ru$^{5+}$ and Ru$^{4+}$  results in a lower theoretical value of  $\mu_{eff}$ given by $\sqrt{0.86\mu_{Ru^{5+}}^{2} + 0.14\mu_{Ru^{4+}}^{2}}$ = 3.591 $\mu_B$, where $Ru^{4+}$ assumed to be in $S$=1 state.\par
Two anomalies were observed in $\chi _{M}~vs.~T$ and $\frac{d\chi _{M}T}{dT}~vs.~T$ (FIG. \ref{MT_CSRO}(a)(i)) at $H$ = 1000 Oe, a sharper peak near 40 K and a boarder peak near 25 K, corresponding to two possible magnetic transitions. We also observed a strong hysteresis in the zero-field cooled (ZFC) and field cooled (FC) $\chi_M$  (FIG.\ref{MT_CSRO}(a)(ii)) at a low field of $H$ =  100 Oe, noticeably the splitting starts above the transitions. Such hysteresis is usually observed in systems with domain magnetization having either a short-range ordered  canonical/ cluster spin-glass state or a long-range ordered ferrimagnetic or ferromagnetic or canted antiferromagnetic state\cite{canted_AFM_2,ferromagnetic_ordering, LRO_ferrimagnet_2}. The field-dependent magnetization isotherm ($M$ $vs.$$ H$) measured at 2 K also exhibits a weak hysteresis with no evidence of saturation and a coercive field $H_c$ = -3.06 kOe (FIG.\ref{MT_CSRO}(b)). At 100 K, $M(H)$ follows almost linear behavior with negligible hysteresis, may be due to short-range correlations up to 100 K. In order to investigate possible spin-glass ground state, we measured temperature-dependent dynamic magnetic susceptibility using an ac driving field of 2 Oe at five different frequencies. FIG.\ref{MT_CSRO}(c) shows the temperature-dependent variation of the real part of ac susceptibility (\( \chi' \) $vs.~T$) at different driving frequencies. A clear anomaly at $T$ = 40 K  was seen in  $\chi'(T)$ without any frequency-dependent shift. Therefore, the magnetization isotherms, temperature-dependent dc and ac susceptibility clearly rule out any spin-glass behavior in Ca$_{2}$ScRuO$_{6}$ rather than indicate a long-range ordered ferrimagnetic ground state or a long-range antiferromagnetic state with magnetic clusters.
\subsection{Heat Capacity Measurements}
\begin{figure}[H]
    \centering
    \includegraphics[width=\linewidth]{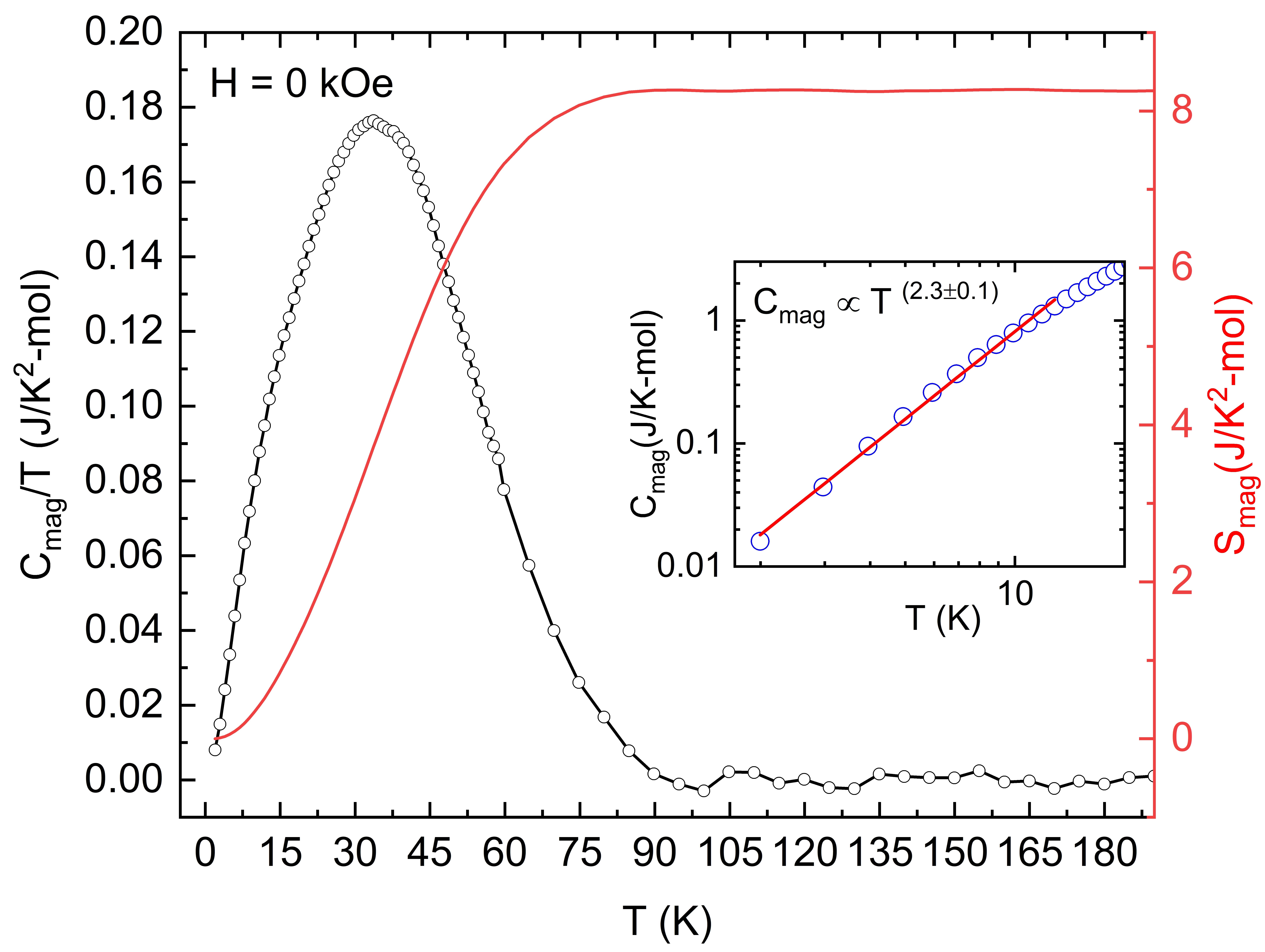}
    \caption{Temperature dependent  heat capacity measurement( $C_p$ $vs.$ $T$) of Ca$_{2}$ScRuO$_{6}$ at an external field $H = 0~kOe$. Inset shows $C_{mag}$ $vs.$ $T$.}
    \label{Cp_CSRO}
\end{figure}
Temperature dependent heat capacity measurements ($C_p$~$vs.$ ~$T$) provide further information on the phase transitions present in the system (FIG.\ref{Cp_CSRO}). No clear anomaly was observed in the $C_p$~$vs.$~$T$ measurement. We extracted the magnetic heat capacity ($C_{mag}$) by subtracting the lattice heat capacity ($C_{lattice}$) from $C_p$. $C_{lattice}$ was obtained by fitting the high temperature $C_{p}~vs.~T$ by the Debye - Einstein model (\ref{App_B}).\par
The $C_{mag}$ shows a broad anomaly around 40 K (inset of FIG.\ref{Cp_CSRO}) which is in line with the anomaly in the magnetization measurements around T = 40 K. This board anomaly suggests that magnetic entropy is spread over a large temperature range up to 100 K, possibly due to presence of small magnetic cluster in a system with weak long-range magnetic order. The corresponding magnetic entropy, ($S_{mag} = \int\frac{C_{mag}}{T}dT$) saturates at 8.26 J/K-mol. For a system with 86\% Ru$^{5+}$ and 14\% Ru$^{4+}$, $S_{mag}$ = 10.719 J/K-mol. The reduction in the  $S_{mag}$ compared to the expected value can result from competing exchange interactions and from overestimation of $C_{lattice}$. Further analysis of low temperature $C_{mag}$ indicated a nearly $T^2$ behavior as shown in the inset of FIG. \ref{Cp_CSRO}, similar behavior was observed in Ba$_2$LaRuO$_6$ and Ca$_2$LaRuO$_6$ \cite{CLRO_BLRO}. 

\subsection{Magnetoelastic coupling} 
\begin{figure}[ht]
    \centering
    \includegraphics[width=\linewidth]{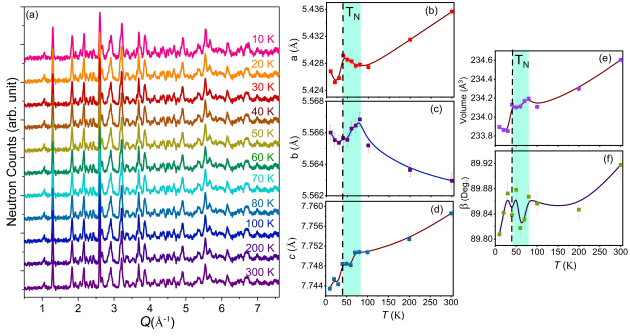}
    \caption{(a) Temperature evolution of neutron diffraction patterns measured from 10K to 300K. (b-f) Thermal variations of lattice constants $a$, $b$, $c$, $V$ and $\beta$ of Ca$_{2}$ScRuO$_{6}$ from 10 to 300 K.}
    \label{NPD_T}
\end{figure}

\begin{figure}[ht]
    \centering
    \includegraphics[width=\linewidth]{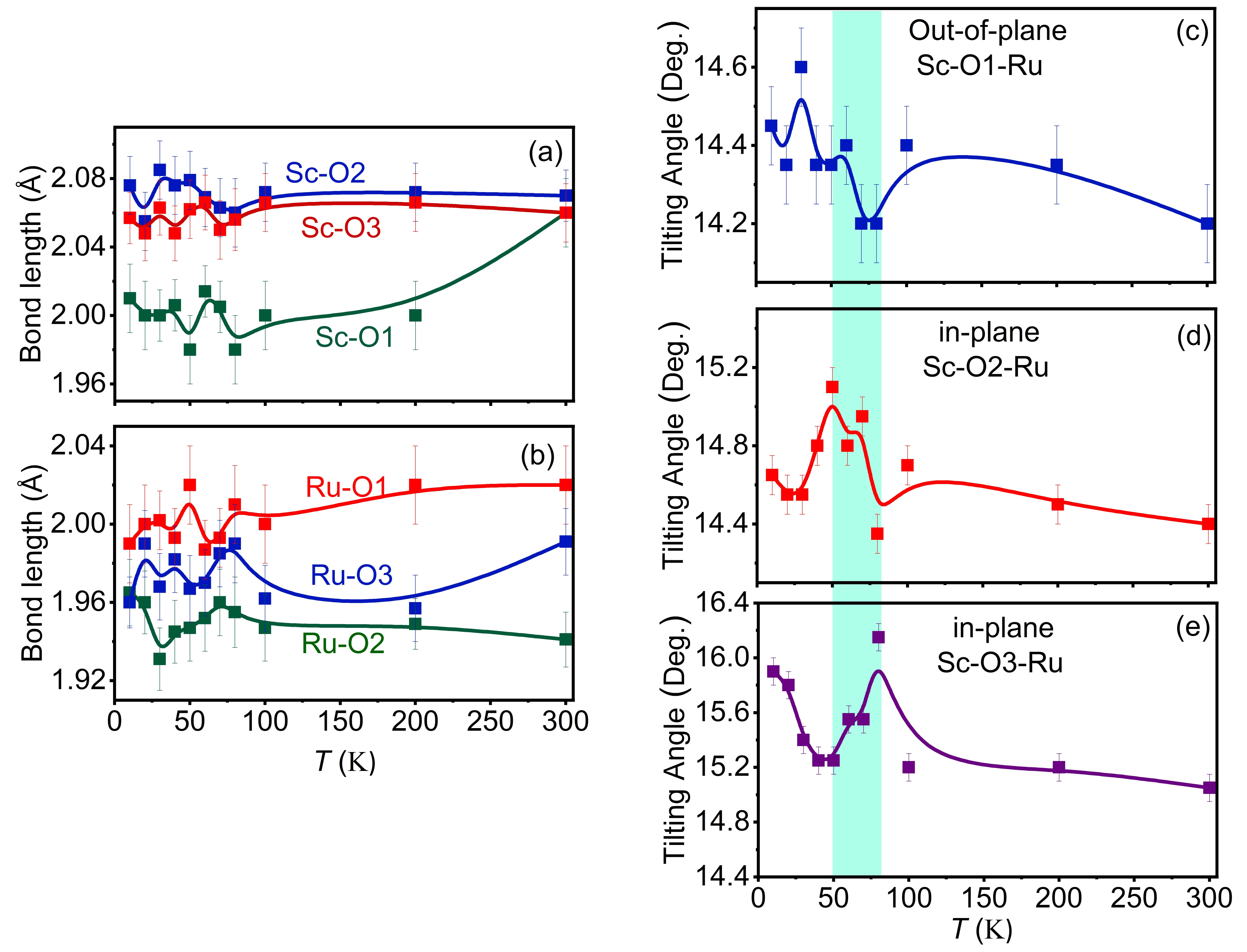}
    \caption{Temperature dependent evolution of bond lengths for (a) $ScO_6$ and (b) $RuO_6$. (c)-(e) show the octahedral tilting angles (Sc-O-Ru) for Ca$_{2}$ScRuO$_{6}$.}
    \label{NPD_bondlength_angle}
\end{figure}

 The temperature dependent NPD patterns also confirmed a monoclinic symmetry (space group $P2_{1}/n$) down to 10 K (FIG.\ref{NPD_T}(a)). The temperature-dependent evolution of the refined lattice parameters showed a significant anomaly around the magnetic ordering temperature of 40 K (FIG.\ref{NPD_T}(b-f)) which reveals a magneto-structural coupling \cite{patel2025crystal,singh2024antiferromagnetic, bera2017zigzag}. Although an overall positive thermal expansion for the lattice constants $a$ and $c$ is evident, the lattice parameter $b$ shows a negative thermal expansion (FIG.\ref{NPD_T}(b-f)). The unit cell volume $V$ also shows an overall positive thermal expansion. The tilting angles~$vs.$~temperature curves (FIG. \ref{NPD_bondlength_angle}(c-e)) also shows anomalies around the magnetic ordering temperature $\sim$ 40 K, revealing a magneto-structural coupling.\par
\subsection{Magnetic order and magnetic structure}
\begin{figure}[ht]
    \centering
    \includegraphics[width=\linewidth]{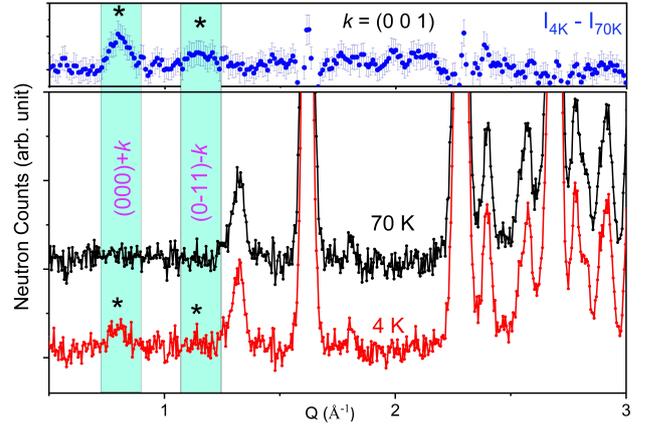}
    \caption{Upper panel: Difference between the diffraction patterns at 4 K and 70 K. Lower
    panel: Temperature evolution of neutron diffraction patterns measured for 4K and 70 K. ${\ast}$ represent magnetic peak.}
    \label{NPD_4_70}
\end{figure}

 In order to identify the presence of weak magnetic order, diffraction patterns were measured at 4 and 70 K with extended counting times ($\sim$ 48 hours per pattern) (FIG. \ref{NPD_4_70}). At 4 K, which is well below the magnetic ordering temperature of 40 K, distinct satellite magnetic Bragg peaks appeared at \textit{Q} $\approx$ 0.8 and 1.15 $\AA^{-1}$. These peaks were not present in the pattern measured at 70 K, where the material is in the paramagnetic state. This confirms the presence of long-range magnetic order in the system. The observed Q-positions of the magnetic peaks corresponds to a propagation vector \textbf{k}$_{mag}$ = (0 0 1).\par
\begin{figure}[ht]
    \centering
    \includegraphics[width=\linewidth]{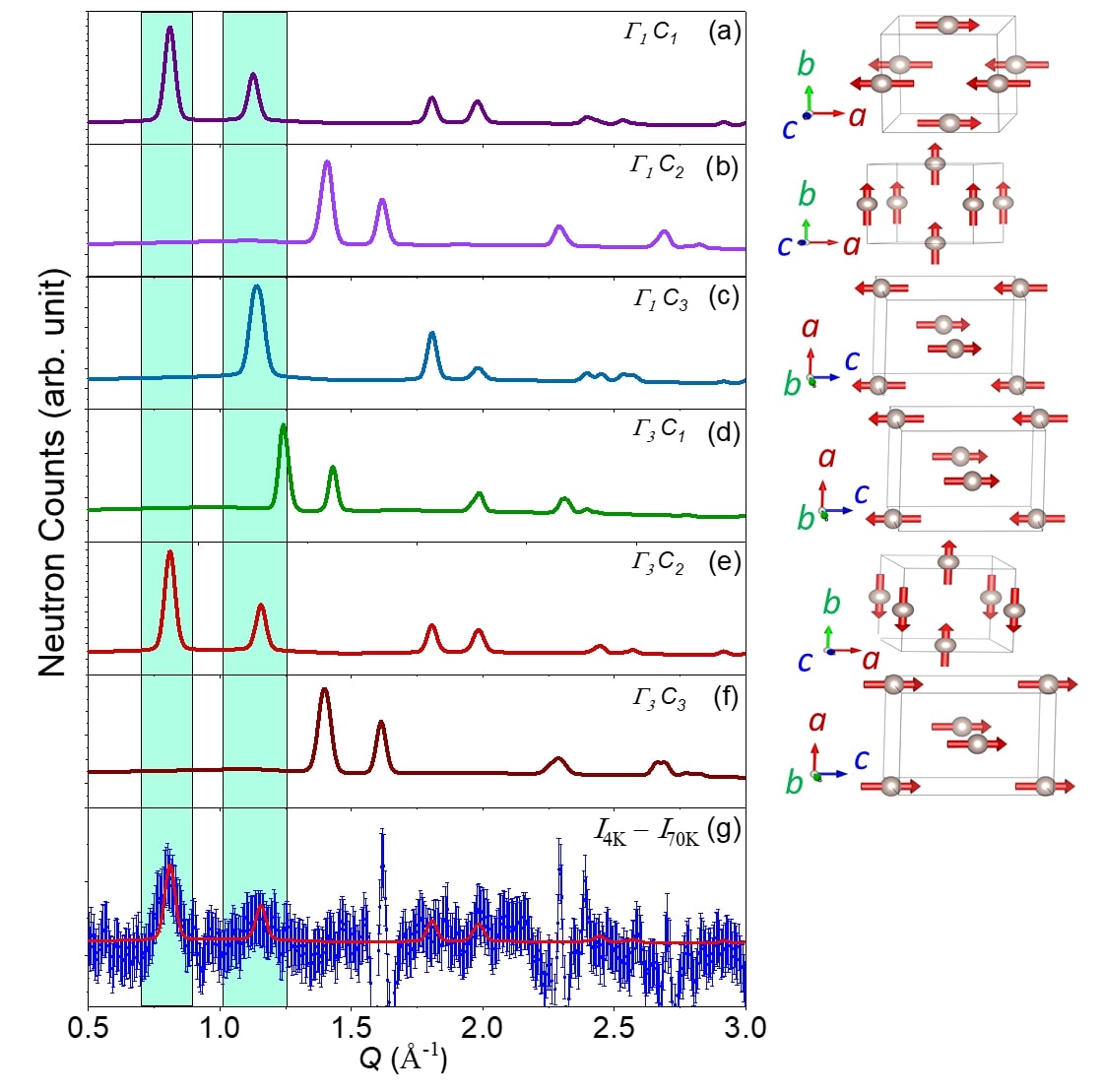}
    \caption{Magnetic refinement patterns for varying gamma values ($\Gamma_1$ and $\Gamma_3$) with $C_1$, $C_2$, and $C_3$ coefficients, showing calculated (line), and difference (bottom) profiles. Here, $C_1$, $C_2$, $C_3$ correspond to the $a, b, c$ components of the magnetic moment. Corresponding magnetic structures are displayed adjacent to each refinement (side panels). Isolation of the magnetic peak via subtraction of the 70 K paramagnetic background from the 4 K dataset}
    \label{mag_structure}
\end{figure}

Symmetry analysis for determining the symmetry-allowed possible magnetic structures was performed using the BASIREPS software within the Fullprof suite\cite{rodriguez1993recent, bera2016long, bera2017zigzag, dutta2022magnetic}. The magnetic Ru$^{5+}$ ions in Ca$_{2}$ScRuO$_{6}$ are located at the $2d$ (0.5, 0, 0.5) site in the space group $P2_{1}/n$. All four symmetry operations of $P2_{1}/n$ leave the propagation vector \textbf{k} invariant. For the propagation vector \textbf{k} = (0 0 1), the irreducible representations, IRs ($\Gamma$) of the little group G$_k$ are given in TABLE \ref{IRs}. There are two possible non-zero IRs which are one dimensional. The magnetic IR $\Gamma^{Ru}_{mag}$ can be decomposed as a direct sum of the IRs as 
$\Gamma^{Ru}_{mag}$ = 3$\Gamma_1$ + 3$\Gamma_3$ \par
The basis vectors represent the Fourier components of the magnetization and are calculated using the projection operator technique in BASIREPS software.TABLE.\ref{basis_vector} presents the basis vectors for the IRs. It indicates that for both IRs, all three magnetic moment components along the crystallographic axes are allowed.\par
The experimental magnetic difference pattern, 4 K-70 K pattern was compared with all possible magnetic structure models defined by basis vectors under the IRs, $\Gamma_1$, and $\Gamma_3$. Due to the weak intensity of magnetic Bragg peaks, a constrained refinement approach was adopted to determine the possible magnetic structure. Each magnetic moment components was systematically varied along with monitoring the appearance of calculated magnetic peaks at expected positions in the diffraction pattern (FIG. \ref{mag_structure}). \par
Three magnetic configurations for the $\Gamma_1$ representation, namely $\Gamma_1C_1$, $\Gamma_1C_2$, and $\Gamma_1C_3$, for three moment components along the $a$, $b$ and $c$ axes, respectively  were investigated. As per the symmetry, $\Gamma_1C_1$-model shows antiferromagnetic coupling between the moments along the $a$ and $c$ axes, while maintaining ferromagnetic alignment along the $b$ axis. For the $\Gamma_1C_2$-model, all moments are aligned parallel way. In $\Gamma_1C_3$-model, the moment components are aligned parallel to each other along the $a$ and $b$ axes and antiparallel way along the $c$ axis. The simulated patterns considering the $\Gamma_1C_1$, $\Gamma_1C_2$, and $\Gamma_1C_3$-models are shown in FIG. \ref{mag_structure}(a-c), respectively (Please note that only the $2d$ magnetic site has been considered for these analyses.). Out of these three models, only the $\Gamma_1C_1$ is in close agreement with the observed magnetic diffraction pattern. Similarly, for the $\Gamma_3$ representation three models $\Gamma_3C_1$, $\Gamma_3C_2$, and $\Gamma_3C_3$ were tested (FIG. \ref{mag_structure}(d-f)) and only the $\Gamma_3C_2$-model provides a close agreement with the observed magnetic diffraction pattern. Therefore, the magnetic structure for Ca$_{2}$ScRuO$_{6}$ can be represented either by $\Gamma_1C_1$ or $\Gamma_3C_2$-model. Both $\Gamma_1C_1$ and $\Gamma_3C_2$-models correspond to A-type antiferromagnetic structure with net zero magnetization per unit cell and moment component along $a$ and $b$ axes, respectively. An ordered moment  of $\sim$1.1(1) $\mu_B$/Ru$^{5+}$ at 4 K was obtained from the refinement by considering one of these models. The ordered moment is significantly reduced from the theoretical ordered moment value of $\sim$2 $\mu_B/Ru^{4+}$ (low-spin $4d^4$; S=1) or 3 $\mu_B/Ru^{5+}$ ($4d^3$; S=3/2). It should also be noted that the small moment components along other directions with a ferromagnetic spin alignment can not be ruled out. The present NPD data set with weak magnetic Bragg peaks does not allow to determine such weak ferromagnetic spin alignment (if any). Single crystal or polarize neutron diffraction data are required for such analysis.\par
\begin{table}[h]
\centering
\caption{Irreducible representations (IRs) of the little group G\textsubscript{k} for the propagation vector \textbf{k} = (0 0 1) and magnetic site $2d$ of Ca$_{2}$ScRuO$_{6}$.}
\begin{tabular}{ccc}
\toprule
\textbf{IR} & \textbf{\{1\,$|$\,000\}} & \textbf{\{2$_{0y0}$\,$|$\,ppp\}} \\
\midrule
$\Gamma_1$ & 1 & 1 \\
$\Gamma_3$ & 1 & -1 \\
\bottomrule
\end{tabular}
\label{IRs}
\end{table}
\begin{table}[htbp]
\centering
\caption{Basis vectors for the Fourier components of magnetization (Ru site) at \textbf{k} = (0 0 1). Space group: $P2_1/n$ (No. 14); Wyckoff position: $2d$. Ru1: (0.5, 0, 0.5), Ru2: (0, 0.5, 0).}
\small
\begin{tabular}{ccccc}
\toprule 
\textbf{IR} & \textbf{Basis Vector} & \textbf{Ru1} & \textbf{Ru2} & \textbf{Magnetic Coupling} \\
\midrule
\multirow{3}{*}{$\Gamma_1$} 
 & $\Psi_1$ & (100)   & (-100)  & \multirow{3}{*}{\parbox{3.5cm}{Antiferromagnetic in $a,c$;\\ Ferromagnetic in $b$}} \\
 & $\Psi_2$ & (010)   & (010)   &  \\
 & $\Psi_3$ & (001)   & (00-1)  &  \\
\midrule
\multirow{3}{*}{$\Gamma_3$} 
 & $\Psi_1$ & (100)   & (100)   & \multirow{3}{*}{\parbox{3.5cm}{Ferromagnetic in $a,c$;\\ Antiferromagnetic in $b$}} \\
 & $\Psi_2$ & (010)   & (0-10)  &  \\
 & $\Psi_3$ & (001)   & (001)   &  \\
\bottomrule
\end{tabular}
\label{basis_vector}
\end{table}
\begin{figure}[ht]
    \centering
    \includegraphics[width=\linewidth]{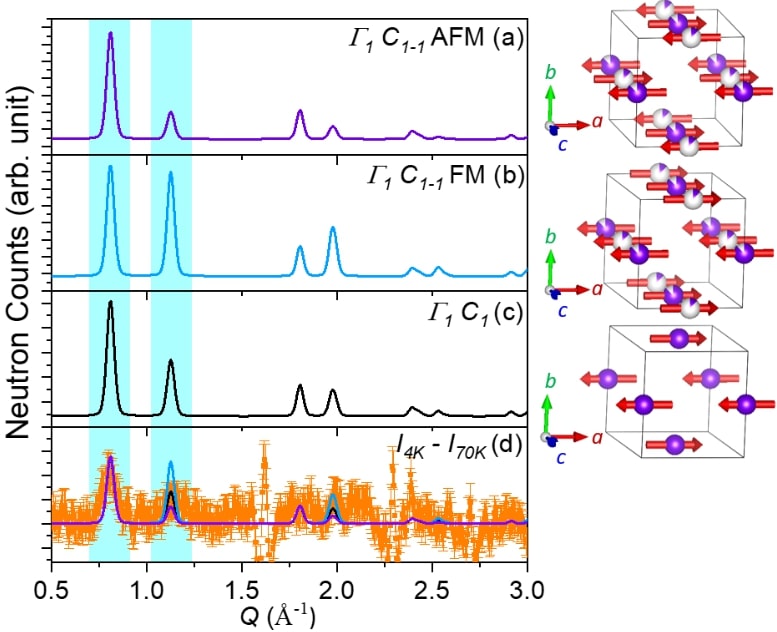}
    \caption{Comparison of magnetic structure refinement models for Ca$_{2}$ScRuO$_{6}$. Simulated magnetic diffraction pattern using (a) a two-site AFM model i.e., magnetic moments for Ru-ions at both the B and B$^{'}$ sites are antiparallel (purple line). (b) two-site FM model i.e., magnetic moments for Ru-ions at both the B and B$^{’}$ sites are parallel (blue line) and (c) a single-site model, i.e., magnetic moment for Ru-ions at the B$^{’}$ site alone (black line) (d) Direct comparison of the single-site (black line), two-site AFM (purple line), and two-site FM (blue line) models with the experimental magnetic diffraction pattern measured at 4 K (orange symbols).}
    \label{mag_structure_comparison}
\end{figure}
In addition, a comparison was done between the simulated magnetic diffraction patterns for two cases: (i) Ru magnetic moments located only at the B$^{\prime}$ site (single-site model) and (ii) antiparallel (two-site AFM model) and parallel  arrangement (two-site FM model)of Ru magnetic moments distributed over both the B and B$^{\prime}$ sites with 14$\%$ and 86$\%$ occupancies  , respectively (FIG.\ref{mag_structure_comparison}). From the direct comparison it was observed that the two-site ferromagnetic (FM) model deviated significantly above the experimental error bars, especially for the peak at \textit{Q} = 1.15 $\AA^{-1}$.The differences between the AFM two-site model  and the single-site model fall entirely within the experimental uncertainties. A detailed analysis of the two-site AFM model is given in \ref{App_C}. However, the single-site model provides the best agreement with the experimental magnetic diffraction pattern with a magnetic moment close to 1.1(1) $\mu_{B}$/Ru$^{5+}$.\par
The NPD refinement results indicate that the magnetic ground state of Ca$_{2}$ScRuO$_{6}$ is a A-type antiferromagnet characterized by a propagation vector \textbf{k} =(0 0 1) and the  magnetic symmetry analysis suggests that a ferrimagnetic structure is unlikely for Ca$_{2}$ScRuO$_{6}$
\subsection{Neutron Depolarization}
Temperature-dependent neutron depolarization (ND) measurements were done to identify the presence of any ferromagnetic moment component in Ca$_{2}$ScRuO$_{6}$. ND technique is highly sensitive to domain magnetization on a mesoscopic length scale ($\sim$100 nm to 10 $\mu$m). It is particularly effective in distinguishing FM, ferrimagnetic, canted-AFM and cluster spin glass states which exhibit finite domain magnetization. In a ND measurement, a polarized neutron beam traverses the sample, its spin precesses around the local domain magnetization, leading to depolarization. This technique thus provides insight into magnetic ordering at the mesoscopic scale \cite{yusef1997polarized,yusuf1996magnetic}.\par
The change in neutron beam polarization was measured for both spin-up and spin-down incident states in Ca$_{2}$ScRuO$_{6}$. The flipping ratio (\textit{R}) between these two polarization states was calculated from the transmitted intensities and is related to the depolarization coefficient (\textit{D}) by the expression \cite{yusuf1996magnetic}:  
\begin{equation}
R = \frac{1 - P_i D P_A}{1 + (2f - 1) P_i D P_A}
\end{equation}
where $P_A$ is the efficiency of the analyzer crystal, $f$ is the efficiency of the DC spin filter, and $P_i$ is the polarization of the incident neutron beam. The transmitted (final) neutron beam polarization ($P_f$) was estimated using the equation, $P_f = DP_i$.\par
\begin{figure}[H]
    \centering
\includegraphics[width=\linewidth]{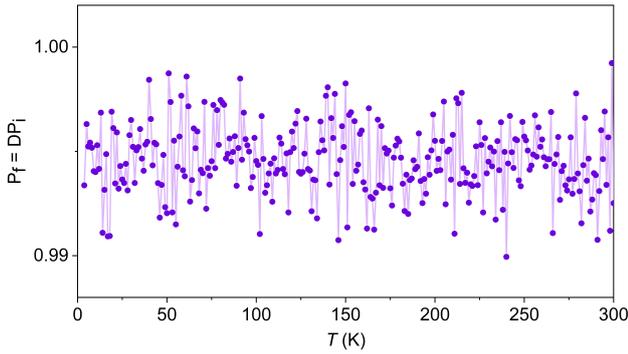}
    \caption{Temperature dependence of transmitted (final) neutron beam polarization ($P_{f}$) over the temperature range of 4-300 K for Ca$_{2}$ScRuO$_{6}$. $P_{f} = DP_{i}$, where $D$ (=1 for paramagnetic state) is the depolarization coefficient and $P_{i}$ is the incident neutron beam polarization.}
    \label{neutron_depolarization}
    \end{figure}

The temperature dependence of $P_f$ over the range of 4–300 K for Ca$_{2}$ScRuO$_{6}$ is shown in FIG. \ref{neutron_depolarization}. The $P_f$ curve exhibits no depolarization, which may be due to the smaller size of the magnetic domains in Ca$_{2}$ScRuO$_{6}$ compared to the detection limit of this  technique ($<$ 100 nm). 
\subsection{Resistivity Measurements}
\begin{figure}[H]
    \centering
    \includegraphics[width=\linewidth]{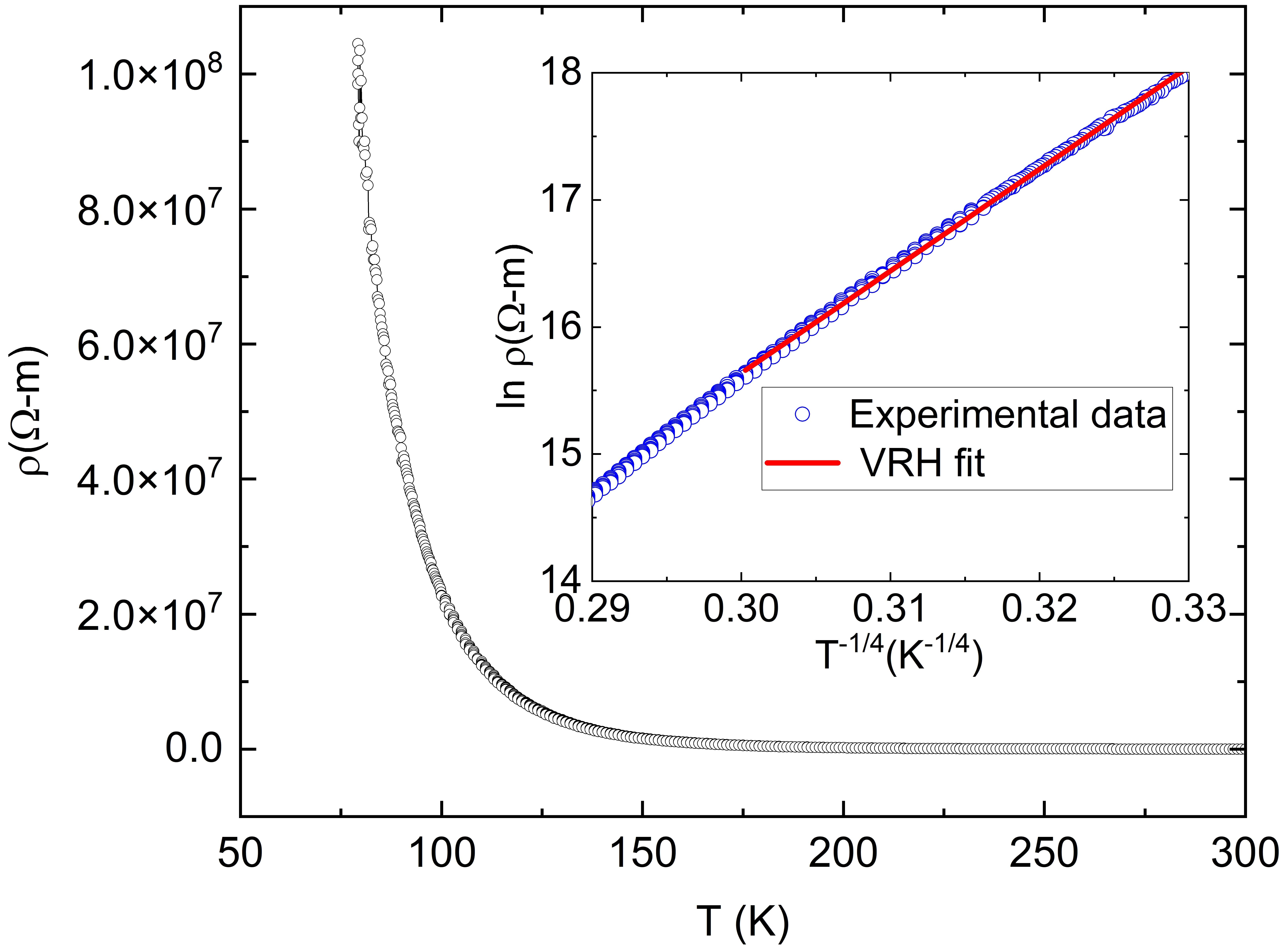}
    \caption{Temperature dependent resistivity measurements ($\rho$  $vs.$ $T$) of Ca$_{2}$ScRuO$_{6}$. Inset shows the 3D VRH fit.}
    \label{rho_CSRO}
\end{figure}
An insulating behavior was observed in the temperature-dependent resistivity measurement ($\rho$ $vs.$ $T$) for Ca$_{2}$ScRuO$_{6}$ (FIG. \ref{rho_CSRO}). In the high temperature regime, $\rho(T)$ followes a 3D Mott- variable range hopping (VRH)  behavior given by $\rho = \rho _{0}$exp$[({T_{0}/T})^{1/d+1}]$(inset FIG.\ref{rho_CSRO}). The dimensionality of the hopping is given by $d$ = 3. The fitted value of $T_0$ is 4.1569 x 10$^{7}$ K. Similar 3D Mott VRH insulating behavior is shown by other polycrystalline samples with localized moments and defect states\cite{3DVRH_Na2IrO3,3DVRH_Li2RhO4,3DVRH_DP_La2CoMnO6}.
\section{Discussions and Conclusions}
PXRD and NPD measurements on polycrystalline Ca$_{2}$ScRuO$_{6}$ confirmed 14\% site mixing between Sc and Ru sites. This resulted in mixed-valency of Ru$^{5+}$ and Ru$^{4+}$ as confirmed from the XAS measurement. Our results clearly indicate that the magnetic ground state of Ca$_{2}$ScRuO$_{6}$ is strongly influenced by the presence of Ru$^{4+}$ state.\par
Two antiferromagnetic-like transitions were observed near 40 K with signatures of domain magnetism. Moreover, a very unusual (for antiferromagnet) hyperbolic inverse susceptibility was observed in the paramagnetic region indicating strong presence of magnetic correlation above the transition temperature. We ruled out any spin-glass ground state from dynamic susceptibility measurement. The puzzle was finally solved only after a careful neutron powder diffraction measurement. The NPD measurement confirmed that the magnetic ground state of Ca$_2$ScRuO$_6$ is a Type-I antiferromagnet with a reduced moment of $\approx$ 1.1(1) $\mu_{B}$/Ru$^{5+}$. In the absence of any antisite disorder, the only possible magnetic ion in A$_{2}$BRuO$_{6}$ ($A$ = Ca-Ba, $B$ = Sc-La) is Ru$^{5+}$, which usually leads to a long range ordered antiferromagnetic ground state with $S$ = 3/2 . 4d$^3$ DP ruthenates such as Ba$_2$ScRuO$_6$, Sr$_2$ScRuO$_6$, Sr$_2$YRuO$_6$, Ba$_{2}$YRuO$_{6}$, Ca$_2$LaRuO$_6$ also exhibit a type-I AFM ground state with an ordered moment between 2.2(2)$\mu_{B}$ and 1.96(2)$\mu_{B}$\cite{SSRO_BSRO,magnetic_structure_Sr2YRuO6_1,2d_magnetic_correlations_Sr2YRuO6,magnetic_structure_Ba2YRuO6,magnetic_structure_CLRO_BLRO}. In these materials, the reduction of ordered moment from 3$\mu_{B}$ ,for $S$ = 3/2 is usually attributed to hybridization between the $4d$ orbital of the transition metal and the $2p$ orbital of oxygen. But in Ca$_{2}$ScRuO$_{6}$ we see that the ordered moment is further reduced. This suggests that a fraction of Ru$^{5+}$ and Ru$^{4+}$ moments do not participate in the long-range magnetic order, rather cause short-range correlation. Spin-glass ordering is ruled out-from ac-susceptibility measurement, and ND measurement does not detect any depolarization of neutrons from magnetic domains. Hence we conclude that Ca$_{2}$ScRuO$_{6}$ has very weak type-I antiferromagnetic ground state in presence of small magnetic clusters ($<$ 100 nm) created by Ru$^{5+}$ and Ru$^{4+}$ moments. This weak magnetic ordering and distribution of magnetic entropy over large temperature range due to the presence of small magnetic clusters is further confirmed from heat capacity measurement. Therefore, a very unique magnetic ordering is present in the synthesized Ca$_{2}$ScRuO$_{6}$ having a combination of AFM order and small magnetic clusters ($<$ 100 nm). Y$_2$CoRuO$_6$, La$_2$NiMnO$_6$, La$_{2}$CoMnO$_{6}$ are examples of other DP TMOs which also exhibit the coexistence of long-range magnetic order with short range interactions.\cite{Y2CoRuO6_dynamicferrimagnetism,La2NiMnO6,LRO_ferrimagnet_2}.
In support of long-range magnetic order, we observe that lattice parameters anomalously changes near the magnetic ordering temperature with overall positive thermal expansion coefficient. This confirms a moderate magneto-elastic coupling in Ca$_{2}$ScRuO$_{6}$ which has been also observed in the similar DP ruthenate, Sr$_{2}$YRuO$_{6}$.\par
In summary, our investigations reveal that the magnetic ground state of Ca$_{2}$ScRuO$_{6}$ is characterized by the coexistence of long-range antiferromagnetic order with small magnetic clusters which results a moderate magneto-elastic coupling in this material. The current work provides a comprehensive picture of the magnetic ground state of the only unexplored DP ruthenate, Ca$_{2}$ScRuO$_{6}$. Furthermore, this article presents some unique features of static magnetic susceptibility in an antiferromagnetic system with small magnetic clusters. It will be very interesting to study the dynamics of ordered and disordered Ruthenium magnetic moment by $\mu$-SR and NMR spectroscopy.

\section*{Acknowledgments}
The authors thank CIF, IIT Palakkad and IIC, IIT Roorkee for the experimental facilities. The authors acknowledge the UGC-DAE Consortium for Scientific Research for XAS measurement facilities and thank Dr. Mukul Gupta and Mr. Rakesh Shah for their support during XAS measurement. SM thanks the DST INSPIRE Faculty grant for research funding. SD acknowledges financial support from the ANRF, erstwhile SERB grant CRG/2022/008740. MAH acknowledges the PMRF grant for providing his fellowship.

\appendix
\section{Structural parameters of Ca$_{2}$ScRuO$_{6}$}
\label{app_A}
\begin{table}[ht]
\caption{The Rietveld refined fractional atomic coordinates, B$_\text{iso}$ and site occupancies for Ca$_2$ScRuO$_6$ at room temperature (300 K) obtained from NPD.}
\resizebox{\columnwidth}{!}{%
\begin{tabular}{ccccccc}
\textbf{Atom} & \textbf{Wyckoff} & \textbf{x/a} & \textbf{y/b} & \textbf{z/c} & \textbf{B$_\text{iso}$ ($\times 10^2$ \AA$^2$)} & \textbf{Occ.} \\
\hline
Ca        & 4e & 0.992(2)  & 0.052(1)  & 0.257(3)  & 0.19(2) & 1 \\
Sc/Ru     & 2b & 0.5       & 0         & 0         & 0.11(4) & 0.86(1)/0.14(1) \\
Ru/Sc     & 2d & 0.5       & 0         & 0.5       & 0.14(2) & 0.86(1)/0.14(1) \\
O1        & 4e & 0.086(2)  & 0.472(1)  & 0.251(3)  & 0.30(4) & 1 \\
O2        & 4e & 0.713(3)  & 0.304(4)  & 0.045(3)  & 0.33(2) & 1 \\
O3        & 4e & 0.194(4)  & 0.209(4)  & 0.952(3)  & 0.30(3) & 1 \\\hline
\end{tabular}}

\textbf{Rp:} 3.60$\%$, \textbf{Rwp:} 4.61$\%$, \textbf{Rexp:} 2.17$\%$, \textbf{$\chi^2$:} 4.53

\label{RR_neutron}
\end{table}
\begin{figure}[ht]
    \centering
    \includegraphics[width=\linewidth]{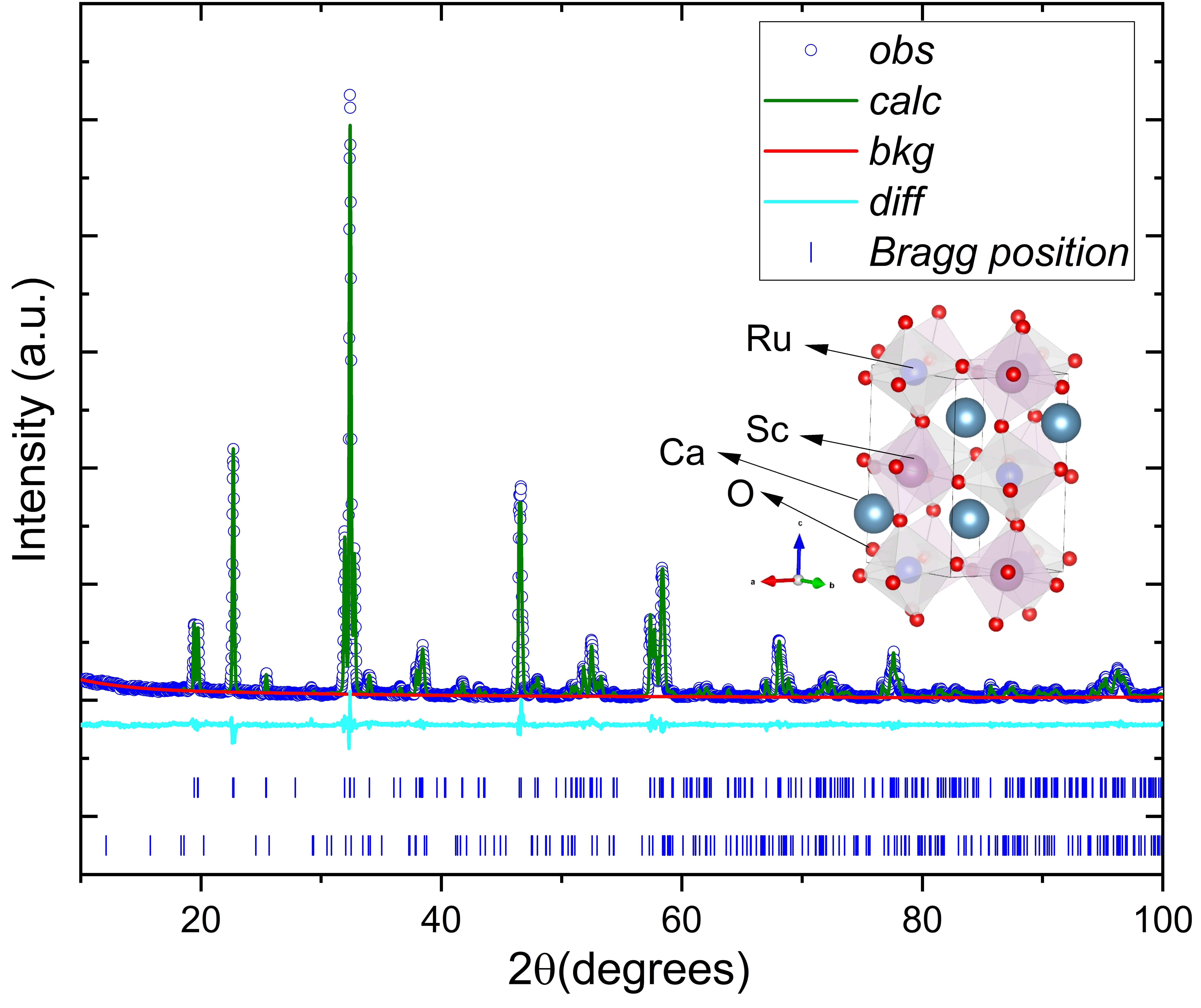}
    \caption{Reitveld refinement of Ca$_{2}$ScRuO$_{6}$ PXRD pattern. The measured data, the calculated pattern from the refinement, the background and the difference between the measured data and the calculated pattern are referred to as \textit{obs}, \textit{calc}, \textit{bkgd} and \textit{diff}, respectively. The vertical blue lines refer to the respective Bragg peak position. The inset shows the crystal Ca$_{2}$ScRuO$_{6}$ crystal structure with corresponding octahedra which is visualized using the VESTA software\cite{vesta}.}
    \label{RR_CSRO}
\end{figure}
\begin{table}[ht]
\centering
\caption{Atomic coordinates of Ca$_{2}$ScRuO$_{6}$ obtained from Rietveld refinement of PXRD data. (FIG. \ref{RR_CSRO})}
\resizebox{\columnwidth}{!}{%
\begin{tabular}{cccccc}
\hline
 Atoms&Wyckoff&fraction&x&y&z\\
      &position& & & & \\\hline
 Ca&4e&1&0.9841(6)&0.0511(3)&0.2480(5)\\ 
 Sc/Ru&2d& 0.886(5)/0.114(5)&0.5&0&0\\ 
 Ru/Sc&2b&0.857(7)/0.143(7)&0.5&0&0.5\\ 
 O1&4e&1&0.093(1)&0.489(1)& 0.249(1)\\
 O2&4e&1&0.705(2)&0.301(2)&0.058(2)\\
 O3&4e&1&0.207(1))&0.216(1)&0.949(2)\\\hline
\end{tabular}}
 \label{wyCLRO}
\end{table}
\begin{table}[ht]
\centering
{\footnotesize
\caption{Refined lattice parameters of Ca$_{2}$ScRuO$_{6}$ at T = 300 K from PXRD.}
\begin{tabular}{ccc}
\hline
Crystal System&& Monoclinic \\
Space group&& \textit{P2$_1$/n} \\
a ($\AA$)&&5.4405(1) \\
b ($\AA$)&&5.57216(9)\\
c ($\AA$)&&7.7627(1)\\
$\beta$ &&89.867(2)\\
$V$($\AA^3$)&&235.329(5)\\
$R_{w}$&& 11.96$\%$\\
\hline
\end{tabular}}
\label{RRlatticeparameters_XRD}
\end{table}
\begin{table}[ht]
\centering
\caption{Bond lengths and bond angles of ScO$_6$ (2b site) and RuO$_6$ (2d site) octahedra, and angles between ScO$_6$ and RuO$_6$ of Ca$_{2}$ScRuO$_{6}$ at room temperature (300 K).}
\small
\resizebox{0.9\columnwidth}{!}{%
\begin{tabular}{l c c c c}
\toprule
\textbf{Category} & \multicolumn{2}{c}{\textbf{Bond length ($\AA$)}} & \multicolumn{2}{c}{\textbf{Bond angle (deg.)}} \\
\addlinespace[3pt] 
\midrule
\multirow{6}{*}{ScO$_6$ (2b site)} 
 & 2$\times$Sc--O1 & 2.06(2) & 2$\times$O1--Sc--O2 & 91.0(1) \\
 & 2$\times$Sc--O2 & 2.05(2) & 2$\times$O1--Sc--O2 & 89.0(1) \\
 & 2$\times$Sc--O3 & 2.02(2) & 2$\times$O1--Sc--O3 & 91.8(1) \\
 &                  &         & 2$\times$O1--Sc--O3 & 88.2(1) \\
 &                  &         & 2$\times$O2--Sc--O3 & 91.5(1) \\
 &                  &         & 2$\times$O2--Sc--O3 & 88.5(1) \\
\midrule
\multirow{6}{*}{RuO$_6$ (2d site)} 
 & 2$\times$Ru--O1 & 1.94(2) & 2$\times$O1--Ru--O2 & 88.1(1) \\
 & 2$\times$Ru--O2 & 1.96(1) & 2$\times$O1--Ru--O2 & 91.9(2) \\
 & 2$\times$Ru--O3 & 1.99(2) & 2$\times$O1--Ru--O3 & 89.8(2) \\
 &                  &         & 2$\times$O1--Ru--O3 & 90.2(1) \\
 &                  &         & 2$\times$O2--Ru--O3 & 89.7(1) \\
 &                  &         & 2$\times$O2--Ru--O3 & 90.3(2) \\
\midrule
\multicolumn{1}{l}{Inter octahedra} 
 &                  &         & 2$\times$Sc--O1--Ru & 151.6(9) \\
 &                  &         & 2$\times$Sc--O2--Ru & 151.2(6) \\
 &                  &         & 2$\times$Sc--O3--Ru & 149.9(7) \\
\bottomrule
\end{tabular}
}
\label{bond_length and bond_angle}
\end{table}
 \section{Debye - Einstein fit of Heat capacity measurements}
 \label{App_B}
 \begin{figure}[H]
    \centering
    \includegraphics[width=\linewidth]{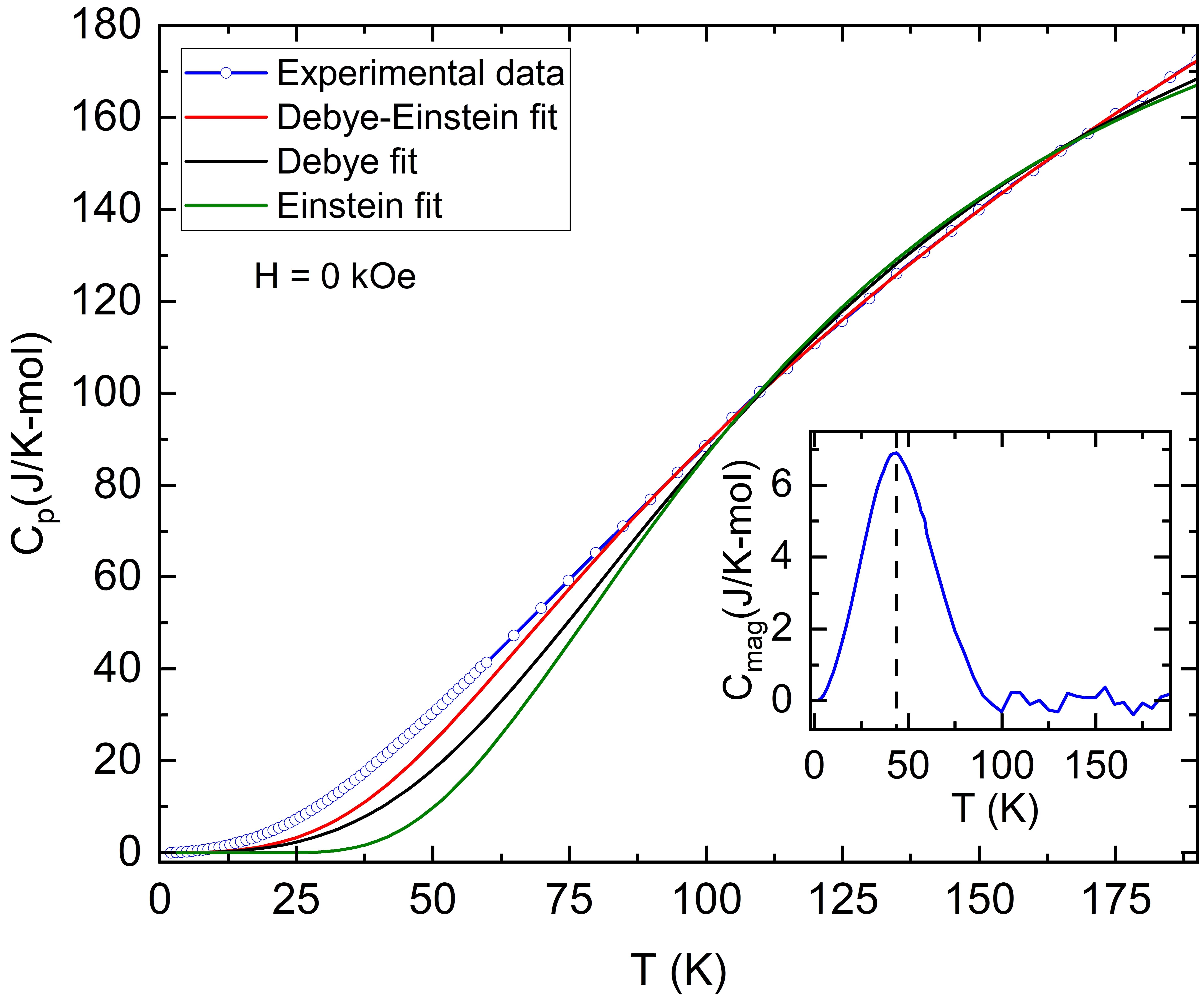}
    \caption{Temperature dependent heat capacity measurement( $C_p$ $vs.$ $T$) of Ca$_{2}$ScRuO$_{6}$ at an external field $H = 0 kOe$. Inset shows corresponding $C_{mag}$ $vs.$ $T$ obtained from the fit.}
    \label{Cp_CSRO}
\end{figure}
The high temperature heat capacity ($C_{p}$) of Ca$_{2}$ScRuO$_{6}$ did not follow either the Debye fit or the Einstein fit ,rather it clearly followed a Debye - Einstein (DE) fit (\ref{Cp_CSRO})  based on the equation \ref{DE_eqn}.
\begin{equation}
\centering
\label{DE_eqn}
  \begin{split}
     C_{lattice}(T) = n_{D}9R\left (  \frac{T}{T_{D}}\right )^{3}\int_{0}^{T_{D}/T}\frac{x^{4}e^{x}}{(e^{x}-1)^2} dx \\
    \\ +  n_{E}3R\left (  \frac{T_{E}}{T}\right )^{2}\frac{e^{T_E/T}}{(e^{T_E/T}-1)^2}
\end{split}
\end{equation}
The first and second term of Eq.\ref{DE_eqn} represent the Debye and Einstein model for heat capacity respectively. Here $n_{D}$, $n_{E}$, $T_E$, $T_D$, $R$, $x$ are Debye coefficient, Einstein coefficient, Einstein temperature, Debye temperature, universal gas constant and $\hbar\omega/k_BT$ respectively. 
 \section{Magnetic structure analysis using two-site model}
 \label{App_C}
\begin{figure*}[t]
    \centering
    \includegraphics[width=0.9\linewidth]{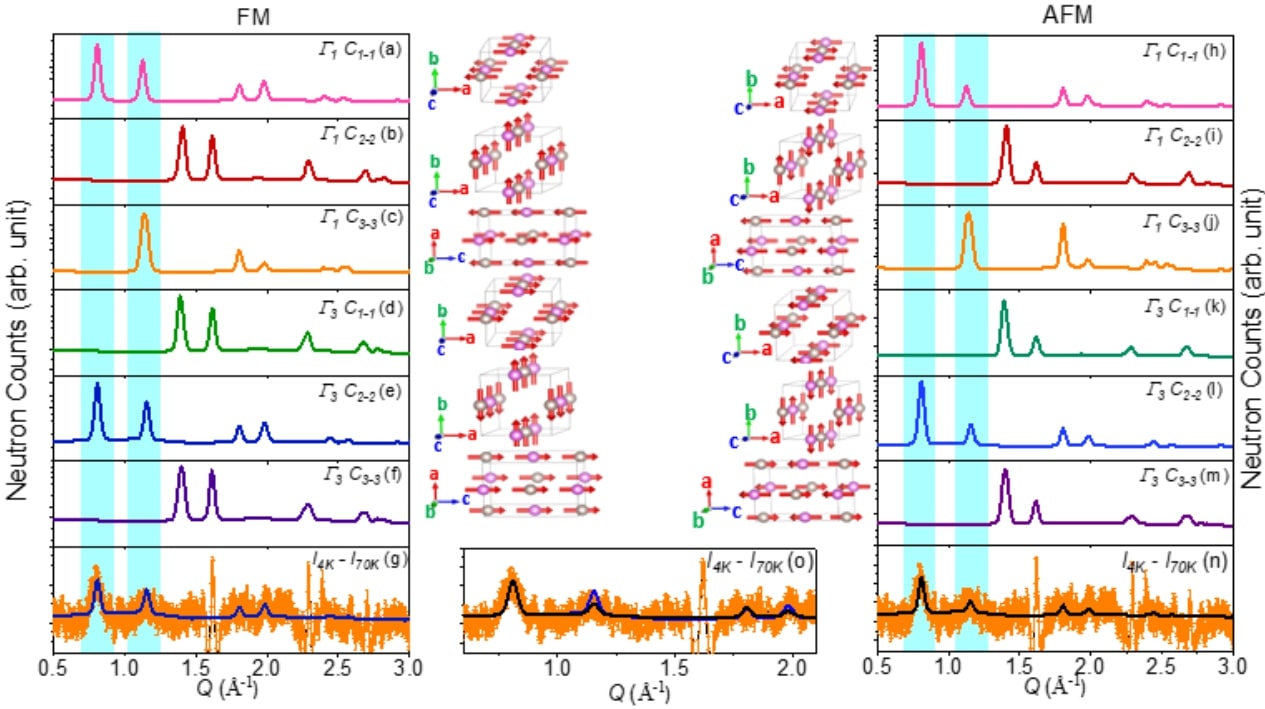}
    \caption{Simulated magnetic diffraction patterns for Ca$_{2}$ScRuO$_{6}$ for the irreducible representations $\Gamma_1$ and $\Gamma_3$ with Basis vectors C$_{1-1}$ (along the $a$ axis), C$_{2-2}$ (along the $b$ axis), and C$_{3-3}$ (along the $c$ axis). Models (a-f) and (h-m) show inter-site ferromagnetic (FM) and antiferromagnetic (AFM) configurations, respectively, between the two symmetry-inequivalent Ru sites B and B’. (g-n) The pure magnetic diffraction pattern (symbols) at 4 K after subtraction of the 70 K paramagnetic background. The best-fit models for each scenario are shown: (g) best fit for FM alignment of Ru moments between B and B’ sites, while (n) best fit for AFM alignment of Ru moments between B and B’ sites (o) comparison between the fits shown in (g) [FM (blue line)] and (n) [AFM (black line)].}
    \label{mag_structure_comparison}
\end{figure*}
 The symmetry analysis for the propagation vector \textbf{k} = (0 0 1) was done for the two-site model using the BASIREPS software. The magnetic representation ($\Gamma_{mag}$) allow to enumerate all symmetry-allowed magnetic structures for both FM and AFM coupling between the Ru moments at the two sites, namely B and B$^{'}$ sites with 14$\%$ and 86$\%$ occupancy.\par
This analysis yielded multiple candidate magnetic configurations. When the spins of both the B ($2d$) and B$^{'}$ ($2b$) sites align in the same direction, the inter-site coupling is classified as FM. These FM models are shown in FIG.\ref{mag_structure_comparison}(a) to (f). For instance, a configuration like $\Gamma_{1}C_{1-1}$ signifies a model where the magnetic moments on both Ru sites are primarily aligned along the $a$ axis. Conversely, the coupling is categorized as AFM when the spins of the two Ru sites align in opposite directions. These AFM configurations between the two sites are shown in FIG.\ref{mag_structure_comparison}(h) to (m).\par
 Among the six possibilities where Ru moments align FM between B and B$^{’}$ sites, the spin configurations shown in FIG.\ref{mag_structure_comparison}(a) and \ref{mag_structure_comparison}(e) produce the magnetic diffraction pattern similar to the experimentally observed one with magnetic Bragg peaks at \textit{Q} $\approx$ 0.8 $\AA^{-1}$ and \textit{Q} $\approx$ 1.15 $\AA^{-1}$.  On the other hand, out of the six possibilities where Ru moments align AFM between B and B’ sites, the spin configurations shown in FIG.\ref{mag_structure_comparison}(h) and \ref{mag_structure_comparison}(l) produce the magnetic diffraction pattern similar to the experimental magnetic diffraction pattern. The magnetic structures shown in FIG. \ref{mag_structure_comparison}(a) and \ref{mag_structure_comparison}(e) are equivalent with different moment directions, i.e., along the $a$ axis and along the $b$axis, respectively. Similarly, the magnetic structures shown in FIG. \ref{mag_structure_comparison}(h) and \ref{mag_structure_comparison}(l) are also equivalent with different moment directions. It may be highlighted that for both these magnetic models, the intra-site spin arrangements (i.e., within the B and B$^{’}$ sites) are AFM. The calculated patterns (FIG. \ref{mag_structure_comparison}(a), (e), (h) and (l)) are shown alongside the experimental pattern in FIG.\ref{mag_structure_comparison}(g) and (n). Finally, a comparison between the inter-site FM and AFM models are given in FIG.\ref{mag_structure_comparison}(o). The fits for both the inter-site FM and AFM cases closely match each other   (FIG.\ref{mag_structure_comparison}(o)). A closer examination of the individual peak intensities reveal that the inter-site AFM model shows a better fit with the data, particularly for the magnetic reflection at \textit{Q} $\approx$ 1.15 $\AA^{-1}$.\par
 Prior to the comprehensive analysis mentioned above, a simpler hypothesis of ferrimagnetic ordering was tested with intra-site FM ordering for both B and B$^{’}$ sites and inter-site AFM ordering (FIG.\ref{mag_structure_comparison}(i) and \ref{mag_structure_comparison}(k)). But these model failed to generate any new magnetic Bragg reflections, even with different moment values considered for B and B$^{’}$ sites. The calculated patterns show only minor adjustments in the intensities of the existing peaks (FIG. \ref{mag_structure_comparison}(i),(k)). Since the simulated pattern was completely different than the experimental one, the magnetic structure is unlikely to be a ferrimagnetic  .
\bibliographystyle{elsarticle-num}

\end{document}